\documentclass[fleqn,usenatbib]{mnras}

\usepackage[T1]{fontenc}
\usepackage{multirow}

\DeclareRobustCommand{\VAN}[3]{#2}
\let\VANthebibliography\thebibliography
\def\thebibliography{\DeclareRobustCommand{\VAN}[3]{##3}\VANthebibliography}

\usepackage{graphicx}	
\usepackage{amsmath}	
\usepackage{amssymb}	
\usepackage{float}
\usepackage[usenames]{color}

\title[Magnetic fields in AGN jets]{Configuration of the global magnetic field in AGN parsec-scale jets}

\author[M.S. Butuzova \& A.B. Pushkarev]{
Marina S. Butuzova,$^{1,2}$\thanks{E-mail: mbutuzova@craocrimea.ru}
Alexander B. Pushkarev $^{1,2}$
\\
$^{1}$Crimean Astrophysical Observatory, Nauchny 298409, Russia\\
$^{2}$Astro Space Center of Lebedev Physical Institute, Profsoyuznaya 84/32, Moscow 117997, Russia\\
}

\date{Accepted XXX. Received YYY; in original form ZZZ}

\pubyear{2023}

\begin{document}
\label{firstpage}
\pagerange{\pageref{firstpage}--\pageref{lastpage}}
\maketitle

\begin{abstract}
The magnetic field plays a significant role in the phenomenon of highly collimated jets of active galactic nuclei (AGN). Relativistic effects prevent the direct reconstruction of the magnetic field direction as transverse to electric vectors on radio maps. We determined the topology of the \textbf{B}-field by modeling the transverse distributions of the total and linearly polarized intensity, polarization degree, and deviation of the polarization direction from the local jet axis and by further comparison with observational data. We consider (i) a helical field with a different twist angle; (ii) a toroidal field on the jet axis surrounded by a sheath with a longitudinal field. In the latter scenario, we consider different sheath thickness relative to the spine. We assumed the sheath velocity is equal to or less than that of the spine. The relativistic effects have been considered for a general case, under which the axis and velocity vector of the jet and radial directions do not coincide. Our simulations reproduce the main features of the observed transverse profiles of polarization characteristics in parsec-scale AGN jets. The model transverse distribution shapes of the polarization properties are found to be strongly influenced by kinematic and geometric parameters of an outflow. We demonstrated it for three AGN having different but typical polarization patterns revealed on radio maps. For each of these objects, we identified the model parameters which provide a qualitative correspondence of theoretical profiles with those obtained from observations, indicating that the \textbf{B}-field is strongly ordered on parsec scales.
\end{abstract}


\section{Introduction}

Active galactic nuclei (AGN) are the most powerful non-transient sources in the Universe. AGN properties are defined by the accretion of matter on a supermassive black hole, one of the appearances of which is producing bipolar outflows. The jets are effectively studied by very long baseline interferometry (VLBI) observations. Developed in the 60s by \citet{Matveenko65}, the VLBI technique enabled achieving record angular resolution at centimetre wavelengths, further improving with the VSOP/HALCA \citep[e.g., ][]{Hirabayashi98, Hirabayashi00a, Hirabayashi00b, Gurvits20} and RadioAstron \citep{Kardashev13, KovalevYuA14, Bruni20, KovalevKardashev20} ground-space interferometers that provided the longest baseline projections. 
On the other hand, an improvement in angular resolution can be reached by increasing the observational frequency. This principle is implemented in the Event Horizon Telescope project \citep[e.g.,][]{EHT14}.
The RadioAstron and EHT projects allow us to investigate the innermost jet regions of active galaxies and improve our understanding of the processes operating there. 
To study the evolution of the morphological structure of the jets, systematic long-term monitoring is needed.
To date, the VLBA program MOJAVE\footnote{\url{https://www.cv.nrao.edu/MOJAVE}} focused on the full Stokes monitoring of bright AGN jets in the northern sky has accumulated the longest-ever observational series. Supplemented by observations made within the framework of its predecessor, the VLBA 2-cm Survey \citep{2cmVLBA,Zensus02}, it, covering the period from 1994 to the present, contains $\approx$450 AGN jets, for each of which there are at least five observational epochs \citep{Lister21}.
Analyzing these data in total intensity $I$, a number of jet properties on parsec-scales have been revealed: (i) velocity distribution of jet features, detection of accelerated motion along curved trajectories \citep[e.g.,][and references therein]{Homan15,Lister21}; (ii) change in the position angle (PA) of the inner jet \citep{Lister13, Lister19, Lister21}; (iii) apparent and intrinsic opening angle and shape of the jets \citep{Pushkarev09, PushkarevKLS17}; (iv) spectral index and brightness temperature, and their change along the outflow \citep{Hovatta14,MOJAVE_XIX}; (v) frequency-dependent synchrotron opacity in the VLBI core \citep{Pushkarev12}.

Magnetic field plays a primary role in the processes of jet formation, acceleration, and collimation \citep[e.g., ][]{BlandfordZnajek77, BlandfordPayne82, Nakamura01, Lovelace02}. 
Its azimuthal component, naturally originated by rotation of accretion disc or black hole, is required to form and then hold the jet. Thus, the helical \textbf{B}-field in the jet is widely expected. 
Based on polarimetric-sensitive VLBA observations at 15~GHz for a sample over 450 sources, \citet{Pushkarev17} found that the linear $\text{PD}=\sqrt{Q^2+U^2}/I$ (where $Q$ and $U$ are the Stokes parameters) typically increases to the jet edges. It was later confirmed by analysis of stacked polarization images for a comparable source sample \citep{MOJAVE_XXI}. The stacked maps showed a much more complete cross-section coverage of a jet in polarization compared to the patchy patterns detected in the single-epoch maps due to the limited sensitivity of the observations. Stacked maps strongly indicate the ordered \textbf{B}-field in jets. Its helical configuration can naturally explain the dip in polarization degree PD) closer to the jet axis due to the partial cancelling of P-signal from regions with different electric vector position angles (EVPAs).

Moreover, \citet{Kharb09, Gabuzda18, Gabuzda21} obtained additional observed evidence of the helical magnetic field by detecting significant gradients of Faraday rotation across the jet. On the other hand, \citet{Laing1980} proposed a model of the spine-sheath structure of the \textbf{B}-field, in which the spine and sheath contain a toroidal and poloidal field, respectively. \citet{Attridge99,Pushkarev05} found supporting this scenario observational evidence in several AGN jets.
Developing this model, \citet{Ghisellini05} suggested that the plasma speed in the sheath is lower than in the spine.
The decrease of the jet flow speed towards the edges was obtained both in the analytical model \citep[e.g.,][]{Beskin17} and analysis of observational data for the nearby active galaxy M87 \citep{MertensLobanov16}.

Therefore, in our simulation of the jet polarization properties, we consider two configurations of the magnetic field: the helical and ``spine-sheath'' structure.
For the latter configuration, we analyse two cases: the sheath speed is equal to or less than in the spine.  
To account for Doppler factor changes caused by a motion of a jet feature along a curved path, we introduce the geometric model in \autoref{sec:jetmodel}. In its framework, the jet component velocity vector, in the general case, does not coincide with the local jet axis and radial direction.
Section~\ref{sec:simul} contains a description of the performed simulation. The results, their discussion, and conclusions are presented in \autoref{sec:res}, \ref{sec:discus}, and \ref{sec:conc}, respectively.

\section{Jet model}
\label{sec:jetmodel}

Typically, VLBI maps of AGN jets show the brightest compact feature, called the VLBI core, and weaker extended regions, tracing the outflow. The core is partially opaque \citep{Hovatta12}, and its position is frequency-dependent \citep[e.g., ][]{Pushkarev12}. 
Downstream from the core, the jet is optically thin.
For simulations, we considered only this case. 
The nature of the bright jet components is still actively debated.
These can be regions with an increased density of radiating particles formed by the central engine initially \citep[e.g., ][]{Stawarz04} or by the development of hydrodynamic instabilities \citep[see, e.g., ][]{Perucho12}. 
Alternatively, the features can be the regions of recollimation or some jet disturbance, for example, a shock wave \citep{Marscher08}.
Also, they may be regions where the jet bends so that the viewing angle decreases, and due to relativistic beaming, the jet radiation increases for the observer.
In addition, it is not known whether the observed features represent the entire jet flow or only some part of it characterised by enhanced emission. Therefore, we analyze the polarization properties transverse to the local jet axis.

\begin{figure}
	\includegraphics[width=\columnwidth]{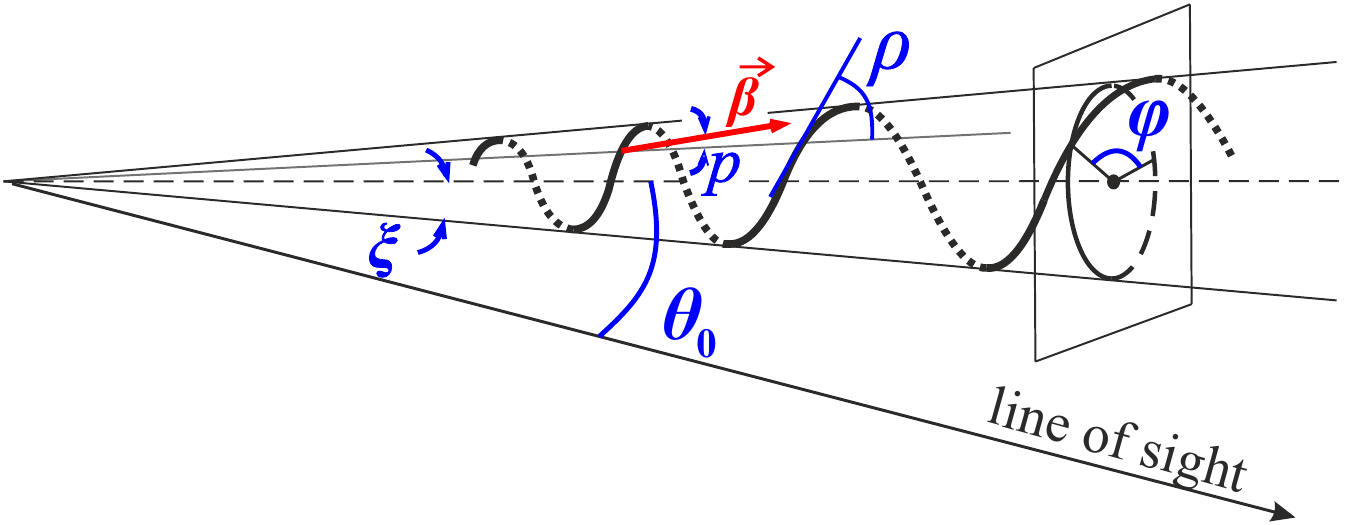}
    \caption{Scheme of the helical jet with designated geometric and kinematic parameters. The thick line denotes the jet. Its parts located on the opposite cone side to the observer are marked by dots.}
    \label{fig:jetmod}
\end{figure}

We use the geometric model of a jet, forming a helix on the surface of an imaginary cone, introduced by \citet{But18a}. Due to the importance of the used geometric parameters for our simulations, we reproduce their description as well as used model parameters and abbreviation in \autoref{tab:param}. 
By the jet segment, we mean the part of the modelled jet formed by two cross-sections, within which the local jet axis can be considered as a straight line.
Adjacent jet segments have different $\varphi$ (\autoref{fig:jetmod}). To maintain the helical shape, the speed of the jet segments must be almost the same. Note that the transition to a particular case --- a straight jet --- can be carried out by setting $\rho=0^\circ$ and an arbitrary constant value of $\varphi$.
If $p=\rho$, the segments move along the jet helix, and we have a constant-in-space jet.
If $p=0^\circ$, the jet segments have radial motion, which manifests through the jet helix's outward motion on the surface of the imaginary cone.
If $p\neq\rho$, during outward motion the jet helix turns around its axis.

\begin{table*}
\caption{Description of model parameters and abbreviations. }
\begin{center}
\begin{tabular}{ll}
\hline
\multicolumn{2}{c}{Model parameters} \\
\hline
 $R_j$ & jet radius \\
 $R$ & distance from the jet axis to the given point \\
 $R_t$ & distance from the jet axis at which the transition from the
spine to sheath occurs\\
$\xi$ & half-opening angle of an imaginary cone \\
$\theta_0$  &  angle of the cone axis with the line of sight \\
$\theta_\rho$ & angle of a local jet axis with the line of sight \\
$\theta_p$ &  angle between a jet segment velocity vector and the line of sight \\
$p$ & angle between a jet segment velocity vector and the cone generatrix at a given point \\
$\rho$ & angle between a jet segment axis and the cone generatrix at a given point \\
$\varphi$ & azimuthal angle of a jet segment  \\
$\beta$ & speed of the jet component or of the spine \\
$\beta_s$ & speed of the jet sheath \\
$\psi^\prime$ & angle of the magnetic field with the jet segment axis (twist angle) in the comoving reference frame\\
$s$ & spectral index of uniform electron distribution $N(E)\propto E^{-s}$, filling the jet segment \\
$I$ & total intensity \\
$P$ & polarized intensity, defined as $\sqrt{Q^2+U^2}$ \\
\hline
\multicolumn{2}{c}{Abbreviation}\\
\hline
PD & polarization degree \\
EV & electric vector \\
PA & position angle \\
EVPA & electric vector position angle \\
\hline
\end{tabular}
\end{center}
\label{tab:param}
\end{table*}

Our model is significantly more complicated than the standard representation of a straight jet moving at a constant angle to the line of sight. Namely, the velocity vector and the local jet axis do not coincide with each other and with the radial direction and, in general, do not lie in the same plane.
Different jet segments have a different angle $\theta$ between the velocity vector and the line of sight because of various $\varphi$s under other constant geometrical parameters of the helix. The change in $\theta$ leads to a change in the Doppler factor $\delta$.
Additionally, for $p\neq0^\circ$, the angle $\theta$ has a wider range of possible values than $\theta_0\pm \xi$ \citep{But18a}. But, only the introduced geometrical model allows to describe self-consistently the following observed facts: 1) the conical shape of the jets on the radio maps stacked over many epochs of observations \citep{PushkarevKLS17}; 2) quasi-periodic changes in the positional angle of the inner (closest to the core) part of a jet detected for more than a dozen sources \citep{Lister13, Lister21}; 3) established both radial and non-radial trajectories of jet features \citep{Lister13, Homan15}.

\section{Simulations}
\label{sec:simul}

\begin{figure*}
	\includegraphics[scale=1]{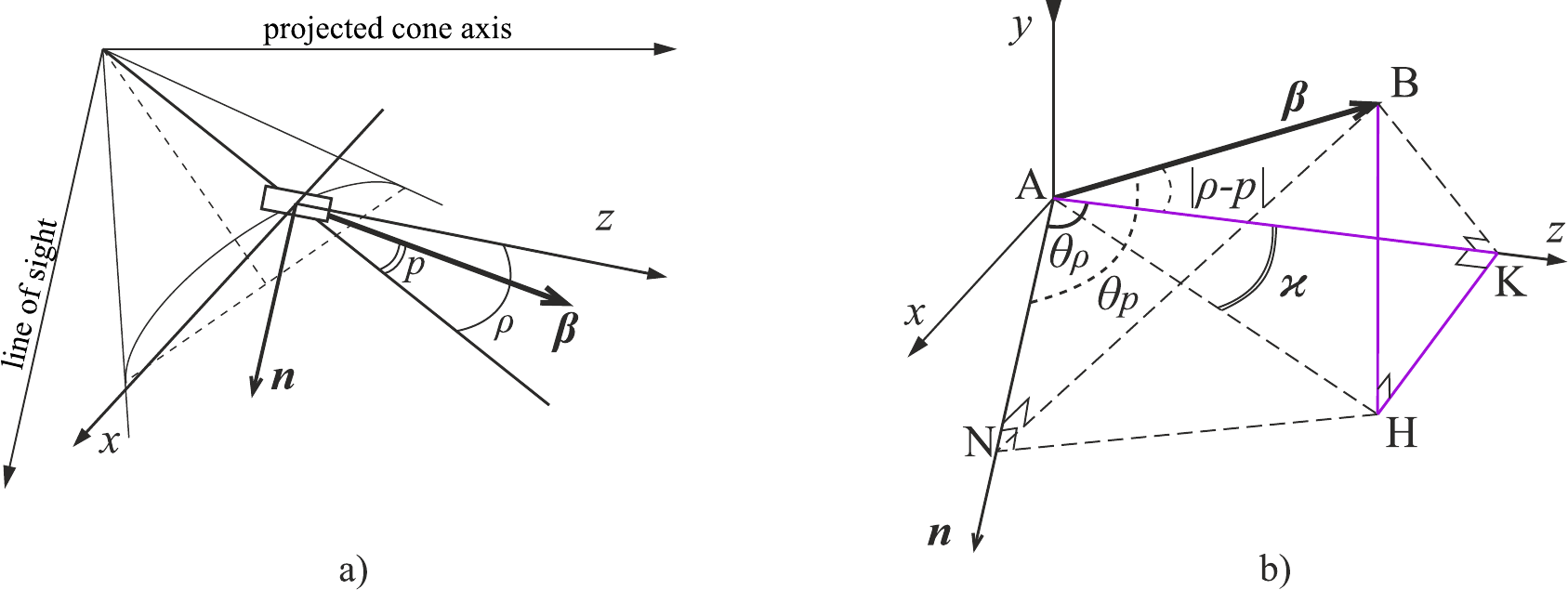}
    \caption{a) The scheme of the coordinate system associating with a separate jet segment with denoted model parameters. The rectangle shows jet segment located on the imaginary cone surface. The cone generatrix, passing through it, makes up the angles of $p$ and $\rho$ with the velocity vector and $z$-axis of the segment (which coincides with a tangent to the jet helix at a given point). The $\boldsymbol{n}$ denotes the line of sight and lies in the $x$--$z$ plane. b) The $\boldsymbol{\beta}$ and its components are highlighted by purple in the used coordinate system.}
    \label{fig:forBeta}
\end{figure*}

To calculate PD and direction of an electric vector (EV) in a wave, we used expressions for the Stokes parameters accounting for the relativistic effects written by \citet{LyutikovPG05} and reproduced in \autoref{sec:appendix}.
Unlike \citet{LyutikovPG05}, we integrated expressions for the Stokes parameters along the line of sight to construct transverse distributions of polarization properties. Previously, including \citet{LyutikovPG05}, two angles of $\theta_\rho$ and $\theta_p$ (see \autoref{tab:param}) were assumed to be the same, and there was no distinction between them.
We consider these two angles separately and allow their values to vary. It is a significant difference from the previously considered models.
The angle $\theta_\rho$ defines the orientation of a rectangular right-hand coordinate system (introduced to specify a \textbf{B}-field in a jet segment) relative to an observer. 
We chose this coordinate system in such a way that the $z$-axis coincides with the local jet axis, and a unit vector  $\boldsymbol{n}$ directed along the line of sight to the observer lies in the $x$-$z$ plane.
The $y$-axis lies in the plane of the sky.
The angle $\theta_p$ is used for Doppler factor calculation. Different jet segments have different values of $\theta_\rho$ and $\theta_p$, which are specified by Eqs. 11-13 in \citep{But18a}
\begin{equation}
    \sin \theta_i \left(i, \xi, \theta_0,\varphi \right)=\sqrt{f_1^2 (i, \xi, \varphi)+f_2^2(i, \xi,\theta_0, \varphi)}\,,
    \label{eq:sinTHpro}
\end{equation}
where
\begin{equation*}
\begin{split}
f_1(i, \xi, \varphi)=\cos i \sin \xi \sin \varphi+\sin i &\cos \varphi \,,\\
f_2(i,\xi, \theta_0, \varphi)= \cos i (\cos \xi \sin \theta_0+& \sin \xi \cos \theta_0 \cos \varphi )- \\
   &-\sin i \cos \theta_0 \sin \varphi\,,
\end{split}
\end{equation*}    
where $i=p$ if we calculate $\theta_p$, $i=\rho$ if $\theta_\rho$ is obtained.
We found components of vector $\boldsymbol{\beta}$ by introducing an auxiliary angle $\varkappa$ and using the scheme displayed in \autoref{fig:forBeta}. The angle $\text{BAK}=|\rho-p|$ because the local jet axis $z$, $\boldsymbol{\beta}$, and cone generatrix, passing through a given segment, lie in the same plane since the distance of the segment from the cone apex is significantly larger than the distance, that the segment passes during the unit time interval.
We found $\beta_x$ and $\beta_z$ from the examination of the ABK and AHK triangles and $\beta_y$ from the BHK and ABK triangles
\begin{equation}
    \begin{split}
     \beta_x=&\beta \cos (\rho-p) \tan \varkappa\,,\\
     \beta_y=&-\beta \sqrt{\sin^2(\rho-p)-\cos^2(\rho-p) \tan^2 \varkappa}\,,\\
     \beta_z=&\beta \cos(\rho-p)\,.
    \end{split}
\end{equation}
We obtained the $\tan \varkappa$  by comparing expressions for AH obtained from the triangles of AHN and AHK
\begin{equation}
    \tan \varkappa=\frac{\cos \theta_p-\cos(\rho-p) \cos \theta_\rho}{\cos(\rho-p) \sin \theta_\rho}\,.
\end{equation}

For numerical simulations of polarization properties, we adopted the value of parameters (\autoref{tab:setnumber}) according to the long-term VLBI observational data for several hundred sources \citep{Pushkarev09,Homan15,PushkarevKLS17, Lister13,Lister16, Lister21}.
The largest uncertainty is in the choice of values of $\rho$. To reduce the free parameters of the model and for the possibility of qualitative comparison of the obtained results, we selected values of $\rho$ that are multiples of $p$ and $\rho<90^\circ$. 

We assumed that magnetic field strength decreases as $1/R$. In the case of the helical \textbf{B}-field in a jet segment, we considered different configurations of the \textbf{B}-field ranging from purely longitudinal ($\psi^\prime=0^\circ$) to toroidal, respectively.
To parameterise the ``spine-sheath'' magnetic field configuration, we used the distance $R_t$, inside of which the toroidal magnetic field is and the poloidal one outside. We have considered various sheath speeds equal to and less than the spine one.
In the simulation, all azimuthal magnetic field components are directed from the observer at positive $y$ and to the observer at negative $y$.
 All parameter values are listed in \autoref{tab:setnumber}.

As a result, we obtained 528 and 990 parameter sets for the helical field and the spine-sheath structure, respectively (see \autoref{tab:setnumber}). For each parameter set, we performed calculations changing $\varphi$ from 1$^\circ$ to 351$^\circ$ in increments of 10$^\circ$.
$\delta$ changes due to changes in $\varphi$ (see \autoref{eq:sinTHpro}) in the interval, which is different for each parameter set.
Additionally, the real jets differ in intrinsic intensity, which decreases with outward distance along jets due to energy losses.
Therefore, to refer to the observed intensity in each particular jet, we divided the interval of $\delta$ changes into three parts with low, intermediate, and high values of $\delta$.
When calculating the Stokes parameters, we integrated along the line of sight at 61 equidistant points on the cross-section of the jet projection on the sky plane. These points locate from $-0.9$ to $0.9$ of the jet radius with a step of 0.03.
To avoid the influence of any edge effects, we considered a part of the jet at distances from the local axis $\leq 0.9$. 
For the adequate comparison of the theoretical and observed distributions of polarization properties, we convolved the former with a one-dimensional Gaussian of FWHM equal to one-third of the jet width.
There are two reasons for one-dimensional convolution. First, transverse distributions smoothly change with a gradual change in $\varphi$. Second, we constructed transverse distributions with a $\varphi$ increment of 10$^\circ$. The distance between jet segments, for which we calculate transverse distribution, is large enough. For example, for a jet length of 10 times its width, a circular two-dimensional Gaussian with the specified FWHM would occupy 1--3 simulated transverse distributions for the whole considered range of $\rho$.
Points on the final simulated transverse distributions of polarization properties, obtained for a given jet segment having corresponding $\delta$, have different colours.
Namely, red, green, and blue refer to the high, intermediate, and low values of $\delta$, respectively, which are achieved with a given set of parameters.

\begin{table*}
\caption{Model parameter variations.  
}
\begin{center}

\begin{tabular}{|c|c|c|c|c|c|c|}
\hline
\multicolumn{6}{|c|}{Jet Geometry and Kinematics} & \multirow{2}{*}{Set numbers}  \\  \cline{1-6}
~  & $\xi$, $^\circ$  & $\theta_0$, $^\circ$  & $p$, $^\circ$ & $\rho$, $^\circ$ & $\beta$ &~  \\
\hline
Linear & 1 & 2, 5, 10 & 0 & 0 & 0.995 & 3 \\
Helical & 1 & 2, 5, 10  & 2, 3, 5, 10 & $\{1, 2, 3, 5, 15, 25\}\cdot p$; $\rho<90^\circ$ & 0.995 & 63 \\
\hline
\multicolumn{6}{|c|}{Magnetic field configuration} & ~\\ 
\hline
Helical & \multicolumn{5}{l|}{$\psi^\prime=0$, 10$^\circ$, 25$^\circ$, 45$^\circ$, 55$^\circ$, 65$^\circ$, 75$^\circ$, 90$^\circ$} & 8 \\ 
Spine-sheath &  \multicolumn{5}{l|}{$R<R_t$, $\psi^\prime=90^\circ$} & 15 \\ 
~ & \multicolumn{5}{l|}{$R>R_t$, $\psi^\prime=0^\circ$}& ~\\ 
~ &  \multicolumn{5}{l|}{$R_t=0.25$, 0.33, 0.5, 0.7, 0.9 of jet radius}  & ~  \\
~ & \multicolumn{5}{l|}{Sheath speed: $\beta_s=0.995$, 0.95, 0.745} & ~ \\
\hline
\end{tabular}
\end{center}
\label{tab:setnumber}
\end{table*}

The final transverse distributions of the polarization properties are the stacked distributions obtained at different values of $\varphi$ for each parameter set.
Therefore, these plots can be compared with distributions obtained from both (i) slices at different distances from the core for single-epoch data and (ii) within a fixed distance interval on stacked multi-epoch maps.
The latter is true, as during the stacked epoch interval, jet parts, characterized by different azimuthal angles, passed through the fixed interval of distances from the core.
Therefore, for comparison with simulated results, we use the transverse profiles of stacked maps constructed with a large enough number ($>20$) of observing epochs and a wide enough time interval ($>15$~yrs) covered by them.

\section{Results and comparison with observational data}
\label{sec:res}

Typically, the observed linear PD shows U-shaped transverse profile, i.e., it is low near the local jet axis and increases towards the jet edges \citep{Pushkarev17, MOJAVE_XXI}. 
In some cases, a W-shaped profile is observed, e.g., in BL~Lac and TXS~1611+343 \citep{MOJAVE_XXI}. The cuts of polarized intensity $P$ have one or, what is striking, two peaks shifted off the jet ridgeline in $I$. 
As for the EV distribution over a source, there are several typical patterns, including predominantly parallel, perpendicular, or so-called ``spine-sheath'' configurations with EV nearly aligned with the local jet direction near the jet axis and transverse at the edge(s) \citep{Attridge99,Gabuzda00,Pushkarev05,ListerHoman05}.

\subsection{Linear jet with radial outward motion}

In our geometrical model, the transition to the linear jet case occurs by setting a fixed $\varphi$. We use the values $\varphi$, at which the jet viewing angle is 2$^\circ$, 5$^\circ$, and 10$^\circ$ (as it follows from \autoref{eq:sinTHpro}). 
Thus, $\varphi=101^\circ$ for $\theta \approx 2^\circ$ and  $\varphi=91^\circ$ for other $\theta$ values.
For radial jet motion $(p=0^\circ)$, the Stokes $U$ is always 0 for any \textbf{B}-field configuration. Therefore, EVs are exactly perpendicular ($Q<0$) or parallel ($Q>0$) to the local jet axis.  

\begin{figure*}
	\includegraphics[scale=0.5]{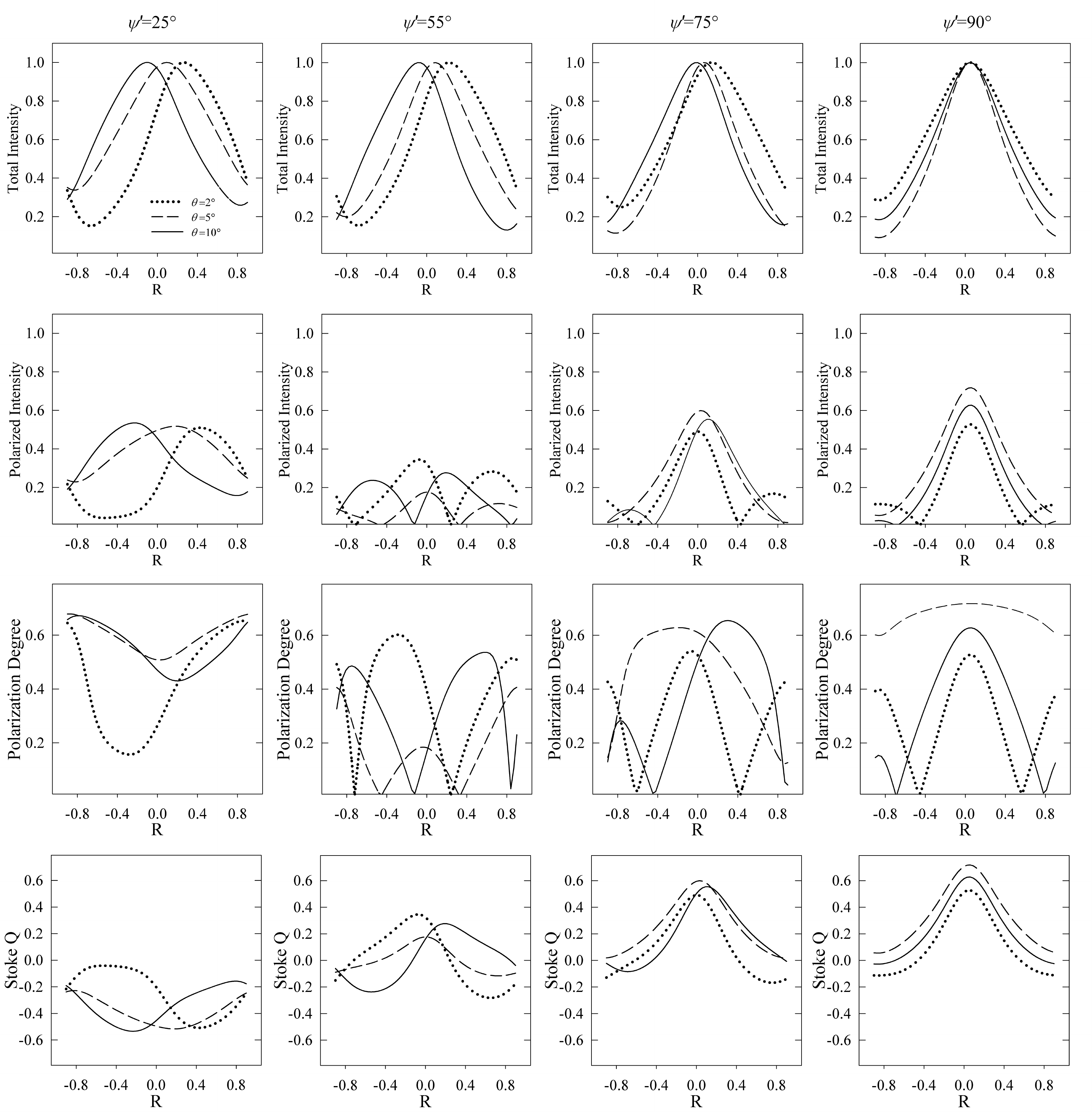}
    \caption{Transverse distributions of $I$, $P$, PD, and the Stokes $Q$ (from top to bottom) for helical magnetic field with $\psi^\prime$ of 25$^\circ$, 55$^\circ$, 75$^\circ$, 90$^\circ$ (from left to right). Dotted, dashed, and solid lines correspond to the jet viewing angle of 2$^\circ$, 5$^\circ$, and 10$^\circ$, respectively.}
    \label{fig:linear_helical}
\end{figure*}

\begin{figure*}
	\includegraphics[scale=0.5]{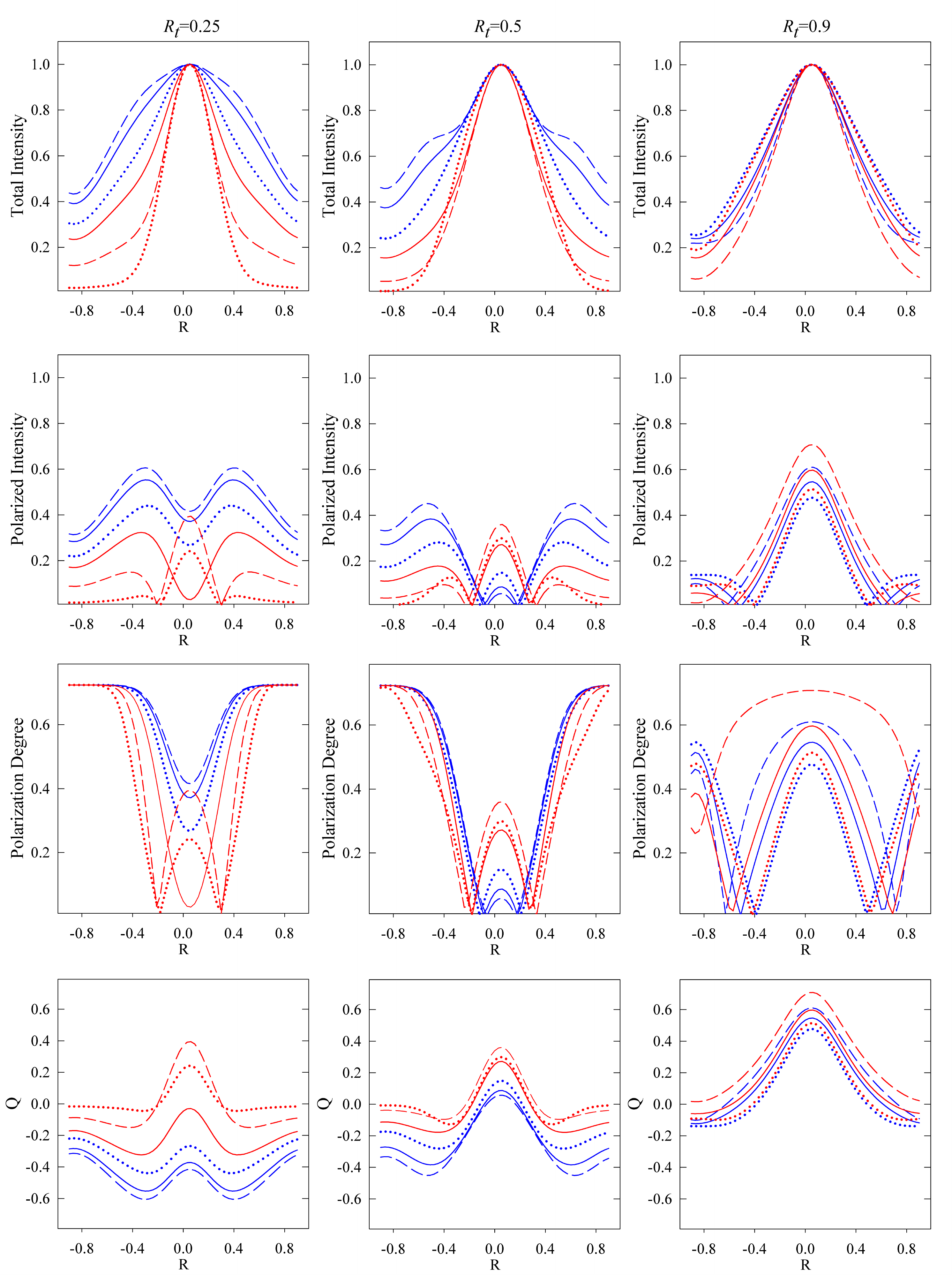}
    \caption{Transverse distributions of $I$, $P$, PD, and the Stokes $Q$ (from top to bottom) for the ``spine-sheath'' magnetic field configuration with $R_t$ of 0.25, 0.5, and 0.9 (from left to right). Dotted, dashed, and solid lines correspond to the jet viewing angle of 2$^\circ$, 5$^\circ$, and 10$^\circ$, respectively. Blue and red lines indicate sheath speed $\beta_s=\beta=0.995$ and $\beta_s=0.745$, respectively. For $Q>0$ or $Q<0$ EVs are parallel or orthogonal to the local jet axis, respectively.}
    \label{fig:linear_spine-sheath}
\end{figure*}

For the longitudinal \textbf{B}-field ($\psi^\prime=0^\circ$), PD reaches its maximum theoretical value \citep[$(\alpha+1)/(\alpha+5/3)\approx0.72$ for $\alpha=(s-1)/2=0.75$, ][]{Pachol}, and its transverse profile is flat. 
The corresponding cuts for $I$ and $P$ have one peak and slight asymmetry. 
With an increase of $\psi^\prime$, a central concavity appears and becomes deeper. \autoref{fig:linear_helical} shows transverse distributions of $I$, $P$, and PD for the different viewing angles and $\psi^\prime\! \geqslant \! 25^\circ$. 
The $P$-distributions have a pronounced asymmetry. 
Only at $\psi^\prime=55^\circ$, this distribution has two-peaked shape with a relatively low level of peaked intensities. Additionally, the PD profiles have a strongly asymmetric W-shape. 
With a further increase in $\psi^\prime$, the $P$-distributions become more symmetrical; their peaks are approximately half the total intensity.
The PD profile has a central peak, which is very high and does not correspond to the observational data of any object.

Let us consider the case of the ``spine-sheath'' \textbf{B}-field configuration (\autoref{fig:linear_spine-sheath}). For a thick sheath($R_t=0.25R$), having a speed equal to that of the spine, the $P$-distribution has two peaks, and there is a concavity in the PD distribution. The Stokes $Q$ is typically negative. The thinner and slower the sheath, the higher value of the Stokes $Q$. As \autoref{fig:linear_spine-sheath} shows, with a decrease in the sheath thickness and/or its speed, the concavity in the fractional polarization profile becomes deeper for cases of $R_t\leq 0.5$.
At the same time, the peak values of$P$ and PD near the local jet axis decrease and then increase. This behaviour is because of the Stokes $Q$ growth  from negative values to zero. For the Stokes $Q$ exceeding zero, the PD and $P$ profiles have the central peak, which increases under further decreasing thickness and speed of sheath. So, the W-shape appearances in PD distributions. 
With a further increase in the Stokes $Q$ for the case of a thin sheath ($R_t=0.9$), the central peak in the PD distribution increases.
It is accompanied by the appearance, growth, and absolute predominance of the central peak in the $P$-profile. In the EV profile, a spine-sheath structure also shows up, namely, inside the jet, EV is directed along the axis, and becomes transverse near the jet edges.

Thus, the simulated profiles of $P$, PD, and EV  agree well with observations for both the helical and ``spine-sheath'' magnetic field configurations. But the model distributions have no point scatter since the jet viewing angle is constant. 
The point spread in the simulated distributions can be obtained by \textbf{B}-field parameter fluctuations or assuming a different degree of magnetic field disordering along a jet. Also, variations of an electron number density and spectral index of the power-law energy distribution of emitting electrons can create the point spread. 
A steady pattern of transverse cuts detected in the observational data \citep{MOJAVE_XXI} indicates that the fine-tuning of parameters necessary to reproduce the observed characteristics in theoretical profiles occurs in at least several jets, which casts doubt on the considered assumption. 
Further, we will show that it is possible to naturally reproduce the observed point spread due to a change in the angle between the jet segment velocity vector with the line of sight when the segments move along curved (helical for complete rotation cycle) trajectories.

\subsection{Non-radial motion in helical jet}

Here we investigate how the transverse distributions of polarization properties change qualitatively with an increase in the twist-angle of the helical magnetic field $\psi^\prime$. 

\subsubsection{Poloidal field}
For an entirely longitudinal \textbf{B}-field, the PD distribution is flat at a value near the maximum theoretical limit. The $P$-distribution has one central peak. The distributions of EV deviations from the local jet axis ($|\text{PA}_\text{jet}-\text{EVPA}|$) are flat with values of near $90^\circ$ or vary throughout the available interval. With an increase in deviation from the poloidal \textbf{B}-field, an asymmetry in all distributions arises and a central concavity appears in the PD distribution (\autoref{fig:heljetnonrad}). 

\subsubsection{Helical field}
\begin{figure*}
\centering
\includegraphics[scale=0.34]{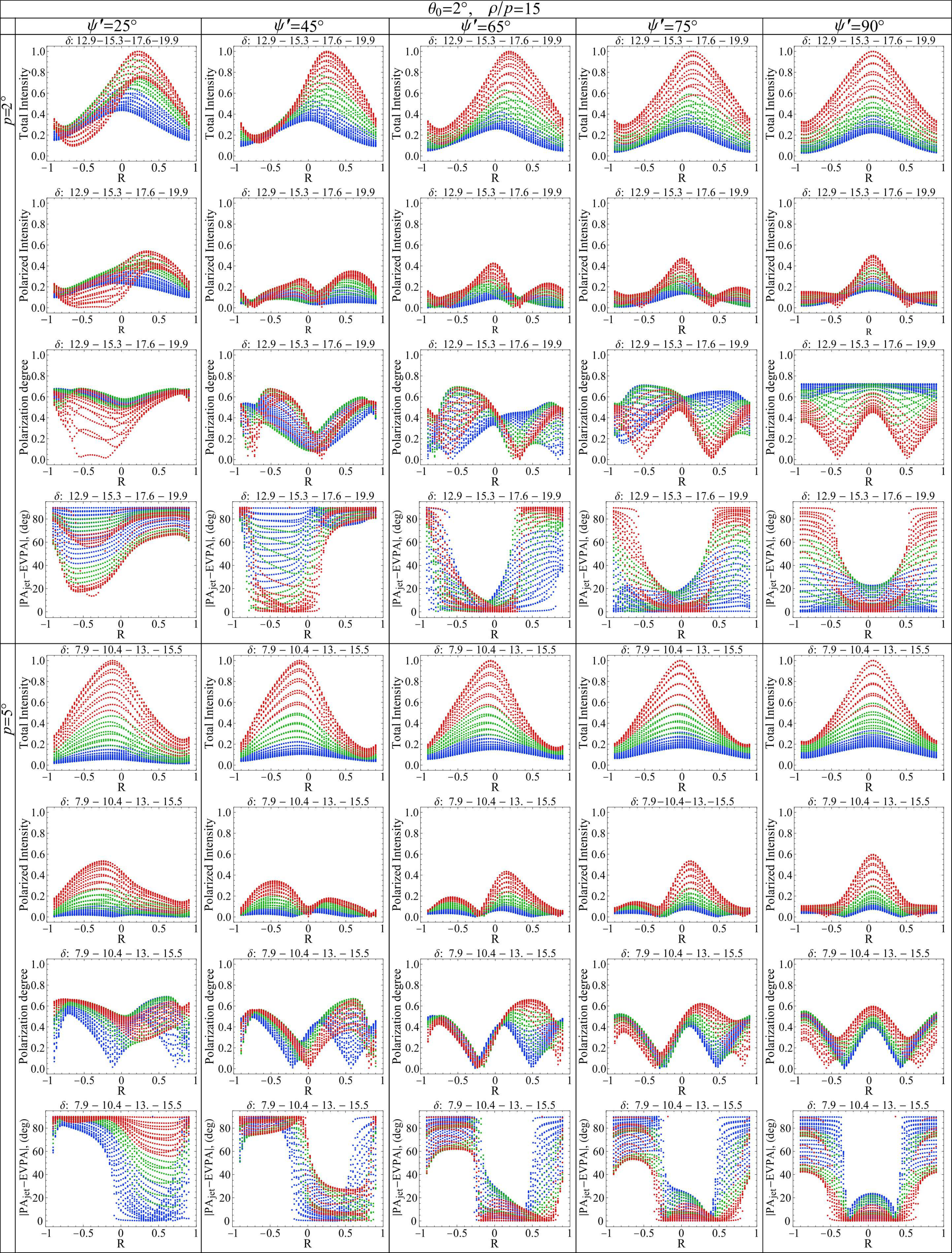}
\caption{Transverse distributions of $I$, $P$, PD, EV deviations from the local jet axis (from top to bottom) for different angles of the magnetic field with the local jet axis $\psi^\prime$. The distributions are given for different angles of the velocity vector of the jet segments with the radial direction: $p=2^\circ$ (upper panel), $p=5^\circ$ (lower panel). Red, green, and blue colours correspond to the high, medium, and low Doppler factor values, lying in the possible range for the given geometrical and kinematic parameters. The Doppler factor intervals are indicated at the top of each model distribution.} 
\label{fig:heljetnonrad}
\end{figure*}

For $\psi^\prime=45^\circ$ and 65$^\circ$, the distributions are strongly asymmetric. The $P$-distribution has two peaks with different magnitudes. 
Not only the direction of the \textbf{B}-field twist but also the values of $p$ determine the dominant peak.
The PD profiles are asymmetric, showing U or W shapes with high peak values. There are three types of $|\text{PA}_\text{jet}-\text{EVPA}|$ distributions: (i) unsystematic spread of points in the entire allowable range of values; (ii) EVs are mainly perpendicular to the jet axis at one edge of the jet and longitudinal on the other; (iii) small values of $|\text{PA}_\text{jet}-\text{EVPA}|$ at the axis and large ones at the edges of the jet for $\psi^\prime\!\!>65^\circ$.
It is important to note that in case (ii) under the same direction of the magnetic field twist-angle, with various other parameters, positive values of $R$ can correspond to both longitudinal and transverse EVs (\autoref{fig:heljetnonrad}, $|\text{PA}_\text{jet}-\text{EVPA}|$ distributions for $\psi^\prime=45^\circ$). 
This fact indicates the ambiguity of determining the direction of the \textbf{B}-field based only on the transverse EV profile, since in model profiles, at $R>0$, the magnetic field lines twirl away from the observer, and at $R<0$, they direct towards the observer.
The skewness sides of distributions of $I$, $P$, and PD for $p=2^\circ$ and $p=5^\circ$ are opposite and cannot help to determine the twirl direction of the magnetic field. The key to solving this problem may be that (1) jet components with different Doppler factors have different profiles; (2) for fixed $\psi^\prime$ and different $p$, the profiles corresponding to low and high values of the Doppler factor are different.

With a further increase of $\psi^\prime$, the asymmetry dilutes.
One peak of $P$ begins to dominate, while the other has a small peak value or disappears altogether in some distributions for $\psi^\prime=65^\circ$. The PD cuts are symmetrical bell-shaped, rarely W-shaped. 

\subsubsection{Toroidal field}
For the toroidal magnetic field $(\psi^\prime=90^\circ)$, all distributions become symmetrical. 
The $P$-profile has one central peak. The PD distribution is mainly bell-shaped with a high maximum value. Predominantly EVs are longitudinal on the axis and transverse to the edges of the jet. The width of the region occupied by the longitudinal EVs depends on both $\theta_0$ and $p$. Sometimes the $|\text{PA}_\text{jet}-\text{EVPA}|$ profiles are mainly within $<20^\circ$ or with an unsystematic spread of points over the entire range of values.

\begin{figure*}
\centering
\includegraphics[scale=0.45]{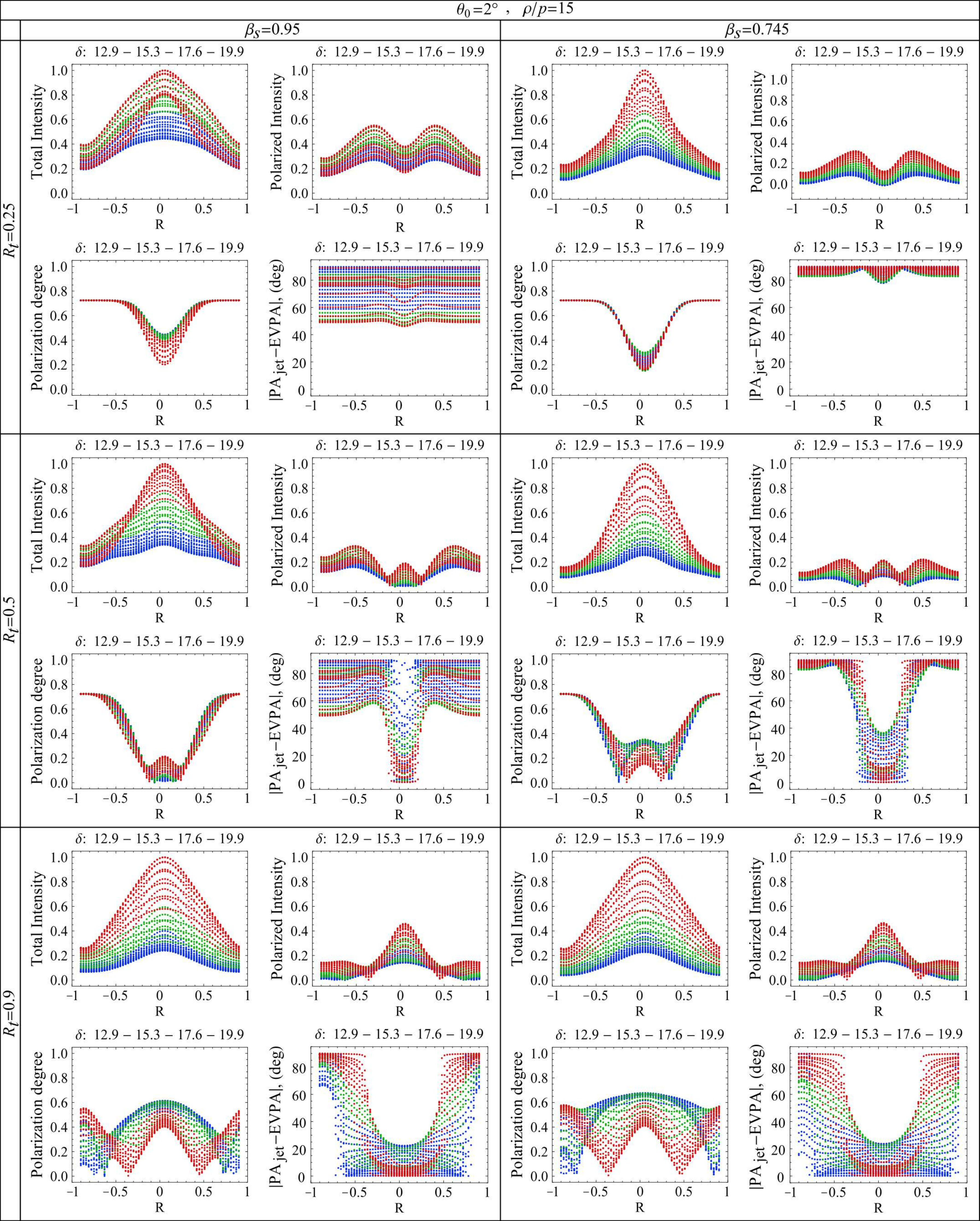}
\caption{Transverse distributions of linear polarization properties in the case of spine-sheath \textbf{B}-field topology. Model parameters are shown. The left and right plot sides correspond to the sheath velocity of $\beta_s=0.995$ (the same as the spine velocity) and 0.745, respectively. The spine radius $R_t$ is 0.25, 0.5, and 0.9 (from top to bottom) of the jet radius. We associate red, green, and blue points with high, medium, and low Doppler factor values corresponding to the given model parameters. The Doppler factor intervals are indicated at the top of each model distribution.}
\label{fig:spine-sheath_heljet}
\end{figure*}

\subsubsection{Spine-sheath configuration}
The ``spine-sheath'' magnetic field structure produces mainly U- and W-shaped transverse PD profiles (\autoref{fig:spine-sheath_heljet}).
The PD profiles are rarely bell-shaped if the sheath is relatively thin. If the sheath and spine velocities are equal, U-shaped PD cuts appear for $R_t\leq 0.33$. 
In this case, the $P$-profile has two peaks equidistant from the jet axis. The EV distributions are flat and almost transverse to the local jet axis, but $|\text{PA}_\text{jet}-\text{EVPA}|$ changes in a wide range, down to 0, for some viewing angles.
For $R_t=0.5$, the W-shaped PD cuts appear for some space of the parameter sets, and the third peak of a relatively small value arises at the jet axis in the $P$-profiles. The ``spine-sheath'' structure in the $|\text{PA}_\text{jet}-\text{EVPA}|$ distribution shows up, in which EVs are longitudinal near the axis and transverse at the edges of the jet. Note  that the jet width with longitudinal EVs is about two times smaller than the width occupied by the toroidal \textbf{B}-field. With a further increase in $R_t$, the central peak values in the $P$ and PD cuts increase. The central peak in the $P$-profiles begins to emerge and dominate in some parameter sets. The width of the region with longitudinal EVs increases too. With further increase in $R_t$, the $P$ and PD cuts become bell- and W-shaped, respectively.
The speed of the sheath penetrated by the longitudinal \textbf{B}-field influences the transverse cuts of the polarization properties. Namely, the slower speed and thinner the sheath, the faster the transition described above from two- to one-peak $P$-profiles, from U-shaped to W-shaped PD cuts, and from flat to spine-sheath distributions of $|\text{PA}_\text{jet}-\text{EVPA}|$. 
We emphasize the particular qualitative profile of, e.g., PD, can correspond to the qualitatively different cuts of other polarization parameters for the given space of the model parameters. For example, the U-shaped PD cut (for $R_t=0.25$ on the top and bottom panels of \autoref{fig:spine-sheath_heljet}) has different profiles of EV deviations from the local jet axis. The W-shaped PD profile for $R_t=0.5$ and $\beta_s=0.995$ and for $R_t=0.9$ and $\beta_s=0.995$ correspond to the $P$-cuts  having two and one peaks, respectively.

\begin{figure*}
\centering
\includegraphics[scale=0.45]{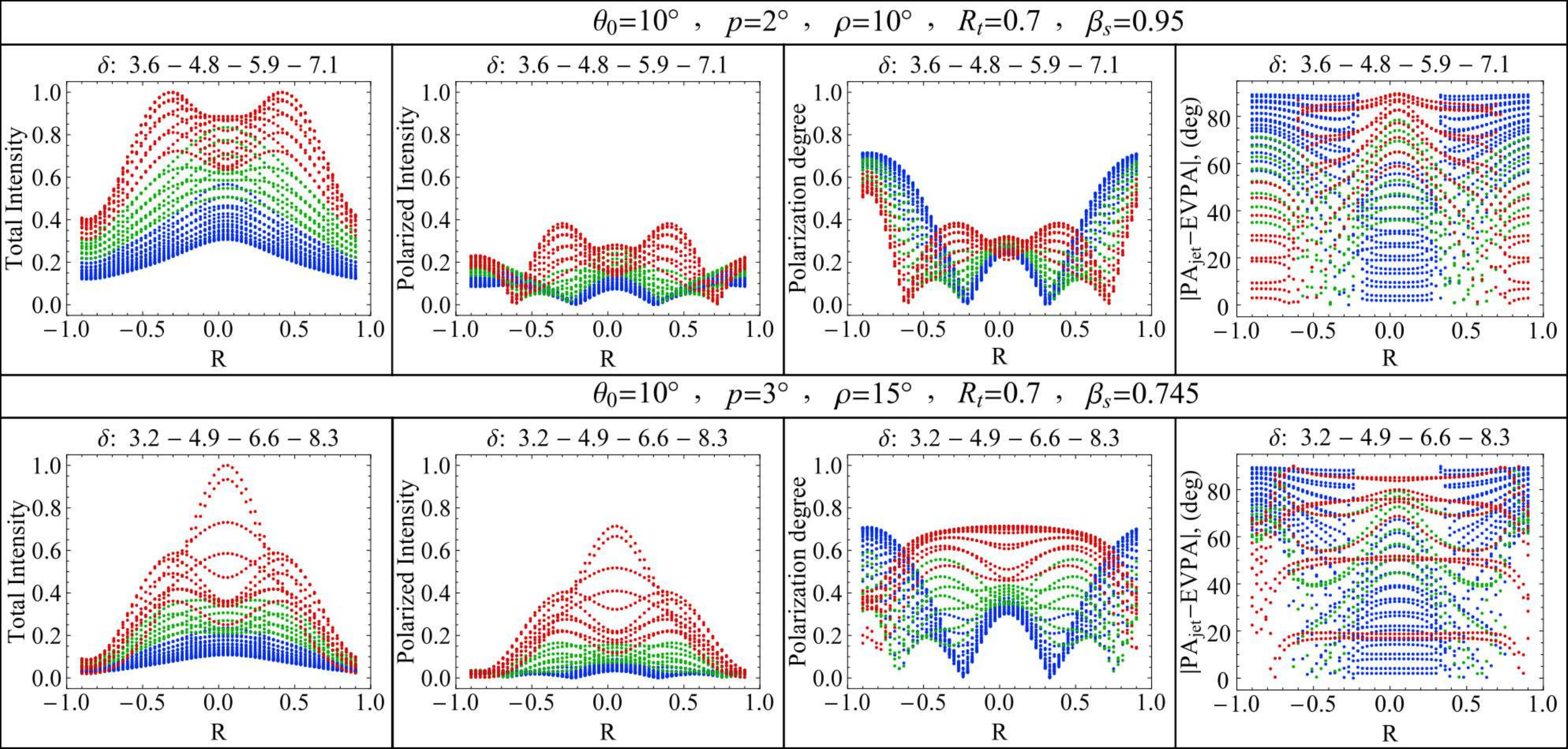}
\caption{Examples of the obtained two-peaked intensity profiles with corresponded transverse distributions of polarization properties. Red, green, and blue colours correspond to high, middle, and low values of the Doppler factor within the permissible range. The Doppler factor intervals are indicated at the top of each model distribution.}
\label{fig:twoIpeaks}
\end{figure*}

A serendipitous finding was the indication of a two-peak total intensity profile as a modeling result. Such cuts we obtained for different sheath speeds and $\rho/p$ values, but only for $\theta_0=10^\circ$ and the spine width $R_t \geq 0.5$. We show examples of the transverse two-peaked $I$-profiles and their corresponding polarization property cuts in \autoref{fig:twoIpeaks}. There are no evident parameters for which two-peaked total intensity distribution arises. Highly likely, with an increase in $\theta_0$, such profiles will occur more often. The search for conditions producing two-peak $I$-profiles seems to be an attractive problem, but this is the subject of future research.
Note that the two-peak transverse $I$-cuts were obtained in the simulation for the reverse pinch field \citep{PriorGourg19} and toroidal \textbf{B}-field \citep{KramerMcD21}. For $\psi^\prime=90^\circ$, we overwhelmingly obtained a bright jet on the axis. There are several sets of parameters at $3\leq \rho/p \leq 15$, for each of which both single-peak and two-peak transverse $I$-profiles are.

\subsection{Comparison with observations}

\begin{figure*}
\includegraphics[scale=0.15]{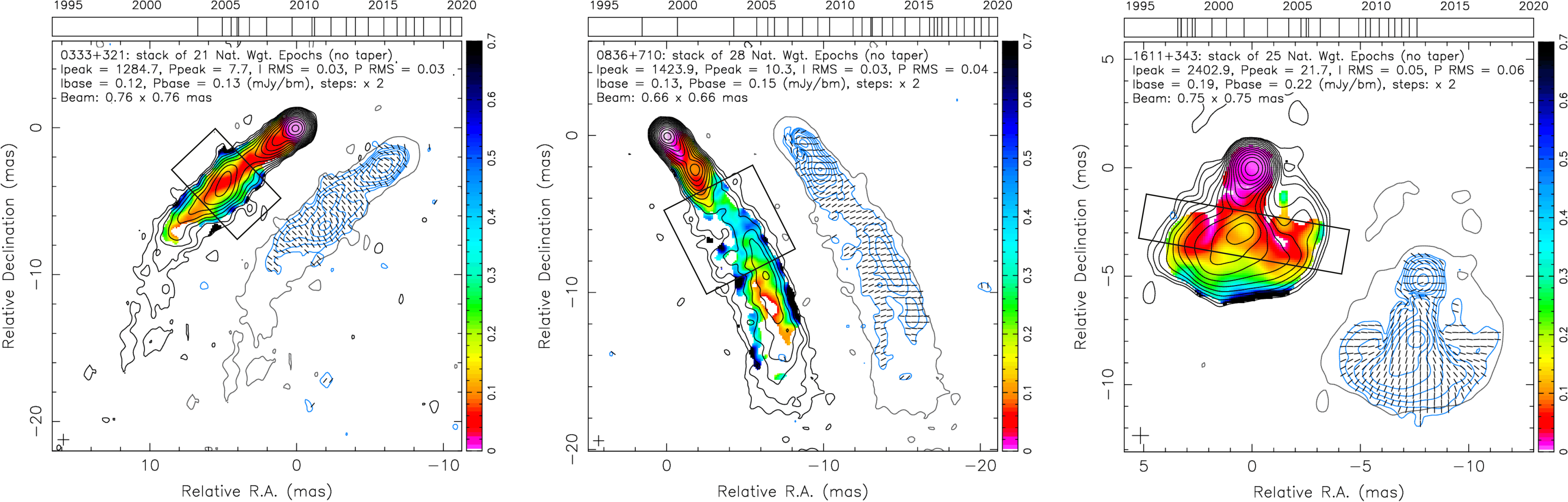}
\caption{Stacked maps of the quasars 0333+321, 0836+710, and 1611+343 (from left to right). Black contours show the total intensity. The linear PD is represented by colour. Blue contours of polarized intensity are constructed relative to the shifted lower one of the total intensity. Sticks display EV directions. Black rectangles indicate the regions, in which we took the observed transverse distributions for further analysis.}
\label{fig:threeobj}
\end{figure*}

\begin{figure*}
\includegraphics[scale=0.3]{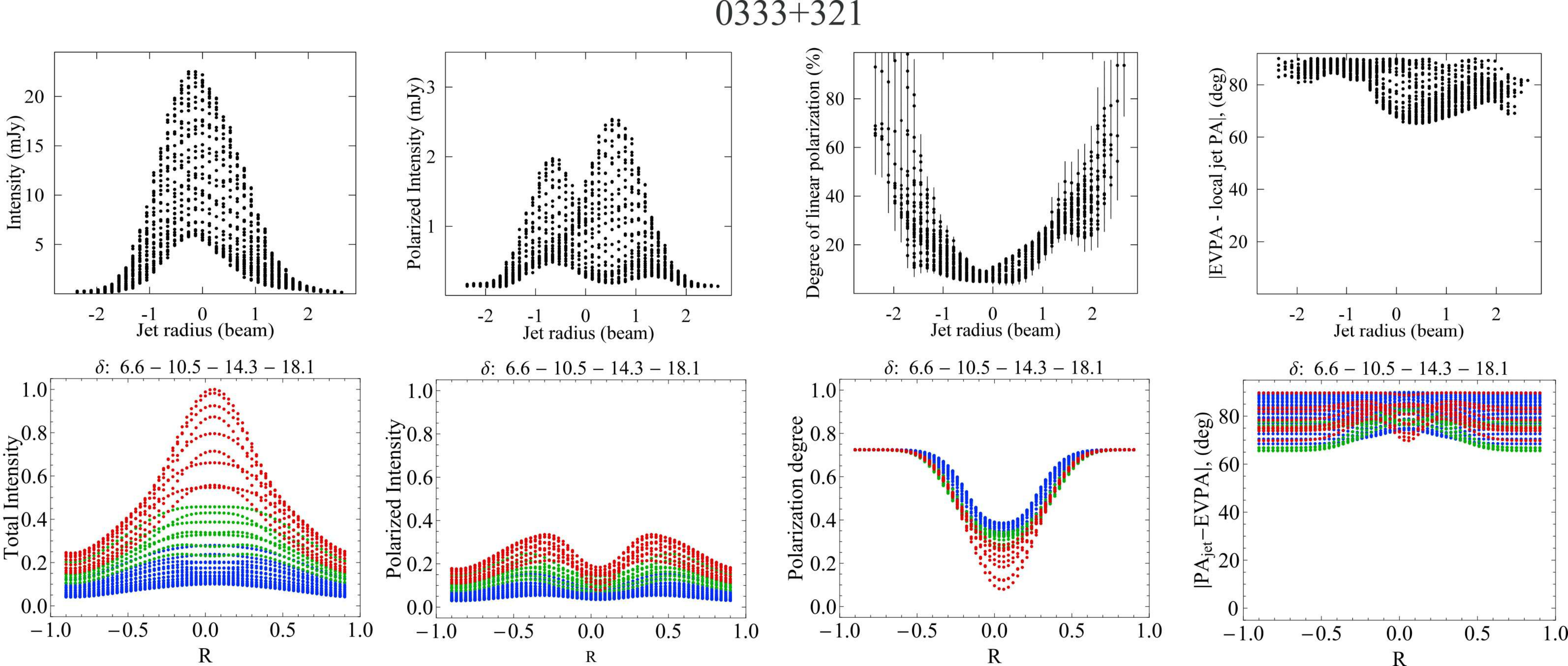}
\caption{Observed (black, top panel) and simulated (colour, bottom panel) transverse distributions of $I$, $P$, PD, and deviations of EVs from the local jet axis (from left to right) for the quasar 0333+321. We took observed data within 4-8~mas from the VLBI core. The model parameters belong to the ``spine-sheath'' magnetic field configuration and are $R_t=0.33$, $\beta_s=0.95$, $\theta_0=5^\circ$, $p=3^\circ$, $\rho=45^\circ$. Red, green, and blue points correspond to the Doppler factor within intervals 14.3-18.1, 10.5-14.3, and 6.6-10.5, respectively.} 
\label{fig:distribs1}
\end{figure*}

\begin{figure*}
\includegraphics[scale=0.3]{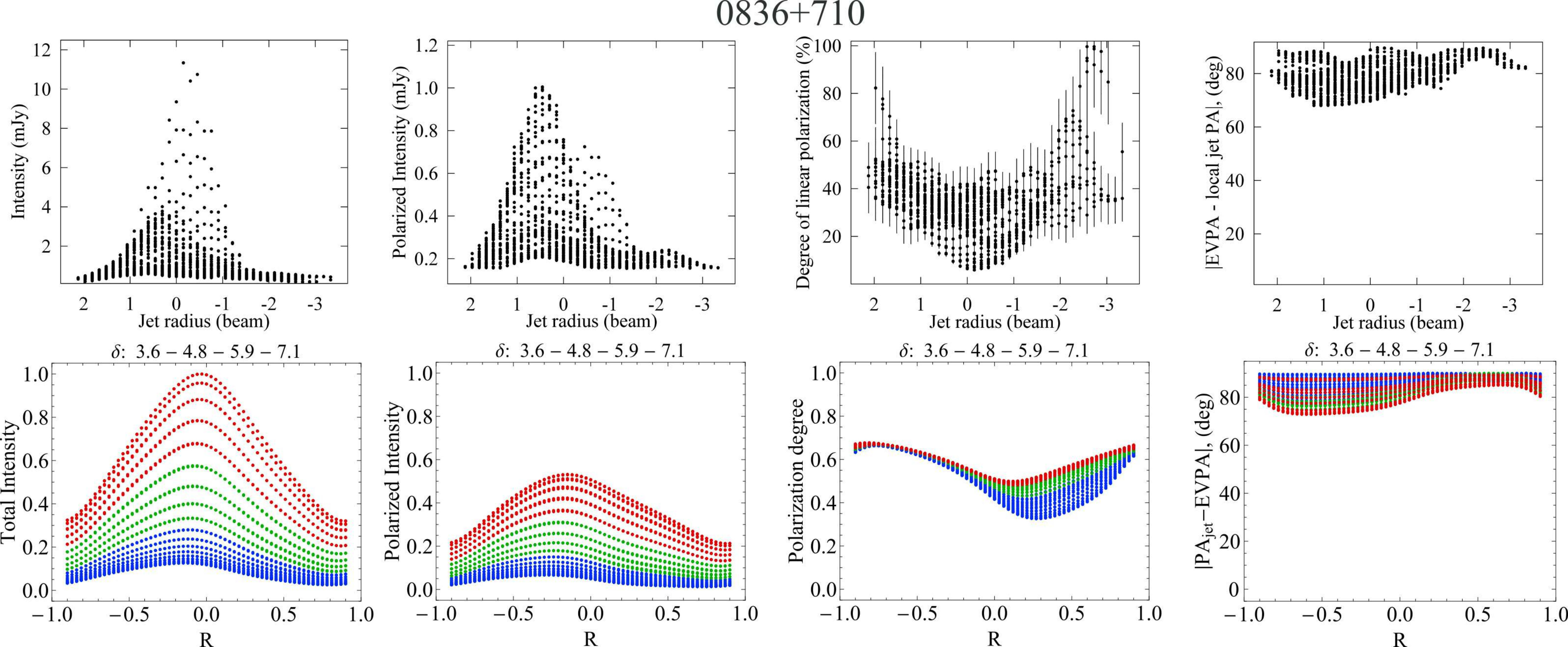}
\caption{Observed (black, top panel) and simulated (colour, bottom panel) transverse distributions of $I$, $P$, PD, and deviations of EVs from the local jet axis (from left to right) for the quasar 0836+710. We took observed data within 4-10~mas from the VLBI core. The helical magnetic field with $\psi=25^\circ$ reproduce the observed data under model parameters: $\theta_0=10^\circ$, $p=2^\circ$, $\rho=4^\circ$. Red, green, and blue points correspond to the Doppler factor within intervals 5.9-7.1, 4.8-5.9, and 3.6-4.8, respectively.
} 
\label{fig:distribs2}
\end{figure*}

\begin{figure*}
\includegraphics[scale=0.3]{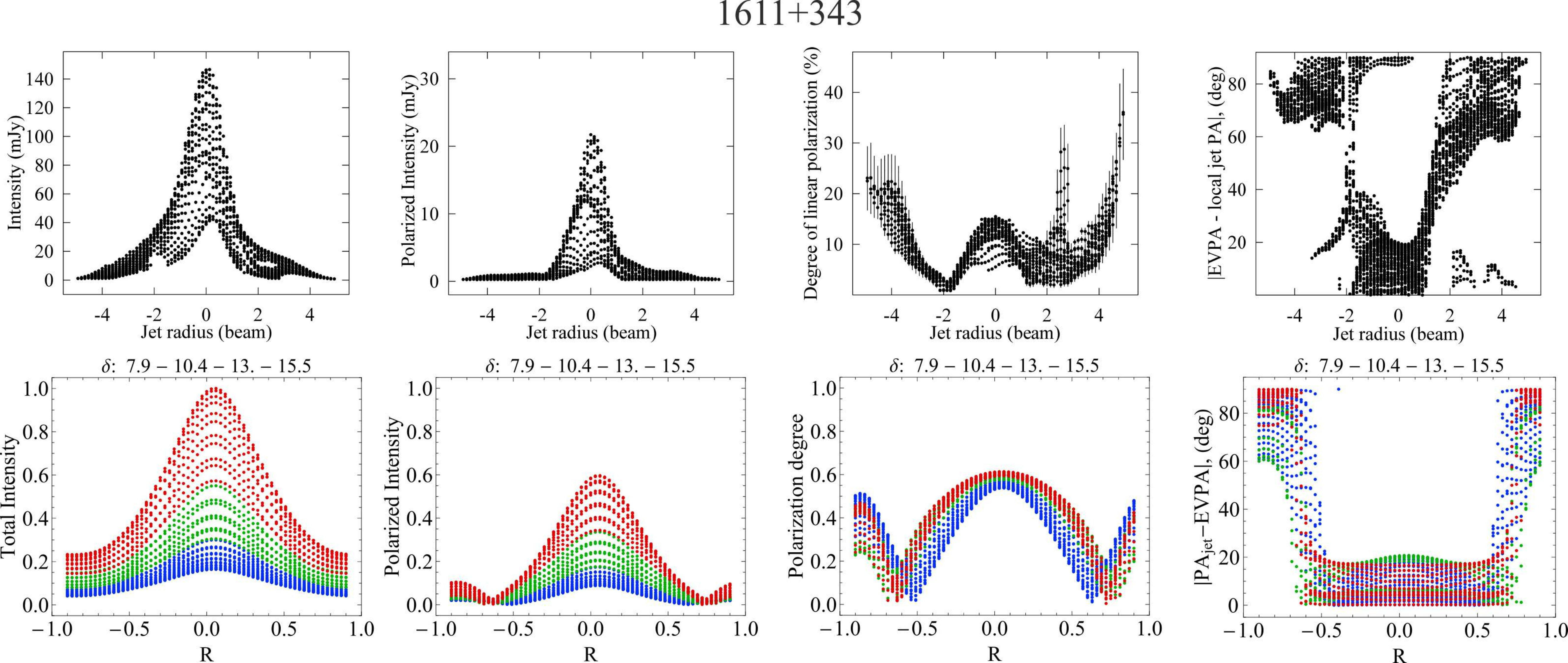}
\caption{Observed (black, top panel) within 2-4~mas from the VLBI core and simulated (colour, bottom panel) transverse distributions of $I$, $P$, PD, and deviations of EVs from the local jet axis (from left to right) for quasar 1611+343. The model parameters belong to the ``spine-sheath'' magnetic field configuration without a difference in the spine and sheath speeds. The model parameters are $R_t=0.9$, $\theta_0=2^\circ$, $p=5^\circ$, $\rho=25^\circ$. Red, green, and blue points correspond to the Doppler factor within intervals 13.0-15.5, 10.4-13.0, and 7.9-10.4, respectively.
} 
\label{fig:distribs3}
\end{figure*}

To compare the simulated distributions of polarization properties with the observed ones, we selected three objects with significantly different transverse PD distributions. 
Since the $\varphi$ of each feature observed on a single-epoch VLBI map is unknown, we used stacked total intensity and linear polarization maps to compare the simulation results with observational data \citep{MOJAVE_XXI}. Thus, we reduced the influence of individual short-term events, and sensitivity increased. Using data in the fixed range of distances from the VLBI core, we are confident that this data covers the entire range of $\varphi$ changes because during the shorter period than considered here, the jet components, on average, completely fill the region with the fixed opening angle on stacked maps \citep{PushkarevKLS17}.
Although, \citet{Hovatta12} obtained that the Faraday rotation of AGN jets from the MOJAVE sample is several degrees and mainly in the region close to the 15~GHz VLBI core, we selected sources with the detected low Faraday rotation to be sure of our conclusions.
\autoref{fig:threeobj} shows the stacked VLBI maps of jets 0333+321 (NRAO~140), 0836+710 (4C~+71.07), 1611+343 (DA~406) in total and polarized intensity. PD and the EV direction are also presented. All maps are taken from \citet{MOJAVE_XXI}.
The black rectangles denote the jet parts used to construct the distributions transverse to the jet ridgeline.
Then, by visual comparison with the simulation results, we searched for the simultaneous correspondence of all four observed distributions with the theoretical ones. We found coincidences for all three considered objects (\autoref{tab:modparobj}, Figures~\ref{fig:distribs1}-\ref{fig:distribs3}). The model parameters corresponding to each source are the only ones in the case of 0836+710 or lie in a narrow interval.

\begin{table*}
\caption{Model parameters that reproduce qualitatively well the character of the observed transverse distributions of polarization properties.}
\label{tab:modparobj}
\begin{tabular}{|c|c|c|c|c|l|}
Object & $z$ & $\theta_0$, ($^\circ$) & $p$, ($^\circ$) & $\rho$, ($^\circ$) & Magnetic field configuration: parameters \\
\hline
0333+321 & 1.3 & 5 & 3 & 45 & ``spine-sheath'': $R_t=0.33$, $\beta_s=0.95$, $\beta=0.995$ \\
~ & ~ & 2 & 2 & 50 & ``spine-sheath'': $R_t=0.33$ or $R_t=0.25$, $\beta_s=0.95$, $\beta=0.995$ \\
~ & ~ & 2 & 2 & 30 & ``spine-sheath'': $R_t=0.25$, $\beta_s=0.745$, $\beta=0.995$ \\
0836+710 & 2.2 & 10 & 2 & 4 & helical: $\psi^\prime=25^\circ$, $\beta=0.995$ \\
1611+343 & 1.4 & 2 & 5 & 25 & ``spine-sheath'': $R_t=0.9$, $\beta_s=\beta=0.995$ \\
~ & ~ & 2 & 10 & 30 & helical: $\psi^\prime=90^\circ$, $\beta=0.995$ \\
\hline
\end{tabular}
\end{table*}

The polarization properties of the 0836+710 jet are well reproduced only by a helical field (\autoref{fig:distribs2}). In this case, even the asymmetry of the theoretical and observed distributions is consistent for all polarization properties.
This fact establishes that the helical \textbf{B}-field directed toward the observer on the brightest side of the jet.
The ``spine-sheath'' structure is unsuitable for this source because, in this case, either a W- or a bell-shaped PD distribution corresponds to one maximum in the $P$-distribution.
Meanwhile, the observed U-shaped PD distribution corresponds to the two-peak $P$-distribution.

On the other hand, only the ``spine-sheath'' \textbf{B}-field structure is suitable for interpreting the observed transverse distributions for the 0333+321 jet (\autoref{fig:distribs1}). 
In the case of the helical field, the obtained from simulations trend of transverse distribution shape changes with $\psi^\prime$ indicates that the two comparable peaks in $P$-distributions are realized for $\psi^\prime=45^\circ-65^\circ$\footnote{Fig.~\ref{fig:heljetnonrad} shows small part of the obtained distributions, all are at \url{ftp://jet.asc.rssi.ru/outgoing/pushkarev/transverse_cuts}}. 
The PD distributions associated with them have a strongly asymmetric W-shape.
Additionally, the variation of EV deviations from the local jet axis occurs in the entire possible range of values. Among the model distributions for the ``spine-sheath'' \textbf{B}-field topology, we found distributions qualitatively corresponding to those observed for 0333+321 in several parameter sets (see \autoref{tab:modparobj}). 

The observed distributions of the quasar 0333+321 exhibit a weak asymmetry, which we can interpret if the field in the spine is not toroidal, but helical with a high value of $\psi^\prime$. Previously, \citet{Asada08} performed detailed VLBI studies the jet in 0333+321 at frequencies 5 and 8~GHz. From the analysis of the EV distribution, the authors concluded that in the radiating region, the \textbf{B}-field direction is at an angle $\gtrsim 80^\circ$ to the local jet axis. In the surrounding jet sheath, containing thermal electrons and thus acting as an external Faraday screen, the \textbf{B}-field inclination is $<10^\circ$. 
If we assume that the considered sheath of \textbf{B}-field extends to the regions with thermal plasma at larger distance from the jet axis, the conclusions of \citet{Asada08} agree well with our simulation results. Moreover, \citet{Asada08} estimate the maximum jet angle with the line of sight at $10\fdg4$ based on kinematics data from \citet{Kellermann04}. Recent kinematic data \citep{Lister21} also confirm this value. From \autoref{eq:sinTHpro} follows for $\theta_0=5^\circ$, $p=3^\circ$, and $\theta_0=2^\circ$, $p=2^\circ$, the angle between the velocity vector of the jet segment and the line of sight does not exceed $8\fdg2$ and $4\fdg2$, respectively.

The transverse distributions for the jet 1611+343 correspond to the case of a thin sheath with a speed equal to spine one. Perhaps, due to the insignificant influence of the sheath, the observed profiles correspond to some theoretical ones for the toroidal magnetic field (\autoref{tab:modparobj}). If we account only for the modelled points corresponding to large values of the Doppler factor, then for 1611+343, there is correspondence at $\psi^\prime=75^\circ$.

Thus, we obtained good qualitative correspondence with the observed data even using a rough parameter grid. Figures~\ref{fig:distribs1}-\ref{fig:distribs3} show that the values at the simulated distributions of PD and $P$ in all cases are slightly larger than at the corresponding observed distributions. As we discuss in the next Section, perhaps, better correspondence can be achieved by varying the model parameters without introducing the disorder of the magnetic field, which, being an additional free parameter, noticeably simplifies the fitting of theoretical distributions to the observed ones.

\section{Discussion}
\label{sec:discus}

Currently, the long-term polarimetric VLBI monitoring of several hundred jets of active galactic nuclei has been performed within the framework of the MOJAVE project at the observing frequency of 15~GHz.
The analysis of these data revealed a tendency for fractional polarization to increase  towards the jet edges, which is present at various distances from the VLBI core \citep{MOJAVE_XXI}.
This finding indicates a well-ordered magnetic field on the probed parsec scales. For example, \citet{Clausen-BrownLyutikov11} showed that the increase in the PD to the jet edge, accompanied by a spectral flattening, is due to a helical \textbf{B}-field.
Therefore, it is necessary to investigate the polarization properties created by a completely ordered magnetic field, the patterns of which cover widely discussed topologies (namely, the helical field and ``spine-sheath'' structure), before concluding about the degree of disorder or the presence of a turbulent component of the \textbf{B}-field.
The latter is often used to interpret sudden jumps in EVPA observed in the optical domain for blazars \citep[for example, ][ for 0836+710]{Raiteri19}.
But \citet{LyutikovKrav17} showed that variations of orientation and velocity of the jet's radiating region could reproduce the observed behaviour of EVPA (smooth swing rotations and sharp jumps by $90^\circ$), accompanied by random changes in PD and $I$ even in the presence of a strictly ordered helical \textbf{B}-field.
Changes in the jet PA and feature speeds are confirmed by long-term observations of several hundred sources carried out within the framework of the MOJAVE project \citep{Lister13,Lister21}.
On the other hand, in the well-ordered \textbf{B}-field, an inhomogeneous distribution of synchrotron photons affects the spectral energy distribution (SED) of both synchrotron and self-Compton radiation \citep{Joshi20}.
Thus, the variations in the SED synchrotron and Compton peak ratio, the frequency shift of these peaks observed during EVPA variations \citep[e.g., the observed ones for blazar 0836+710 ][]{Raiteri19} can be interpreted by a scenario of the well-ordered magnetic field in a jet.

The helical magnetic field is a natural consequence of the jet formation and collimation models \citep[e.g., ][]{BlandfordZnajek77, BlandfordPayne82, Nakamura01, Lovelace02} and explains significant gradients of Faraday rotation measure (RM) across the jets \citep[e.g., ][ and references therein]{Gabuzda21}.
Moreover, if earlier only the presence of different RM signs was a necessary indication of the helical \textbf{B}-field in the surrounding jet environment, then now it is not the case.
Investigating the temporal changes of the transverse RM gradient in the jet 3C~273, \citet{LisakovKrav21} have shown that even the RM with the same sign at different jet edges can indicate the helical \textbf{B}-field in the surrounding jet medium. It occurs as the jet ``highlights'' different parts of the field at different epochs.
But the Faraday rotation occurs outside the synchrotron emitting region, whereas only the the observed EVs can determine magnetic field configuration in a relativistic jet.

For individual jet features from the analysis of single-epoch VLBI observations, \citet{ListerHoman05} identified the preferred directions of EV. Namely, the EV is usually either aligned with or orthogonal to the local jet axis.
\citet{LyutikovPG05} explained this by relativistic effects with an initially helical \textbf{B}-field in the jet reference frame. In their formalism, EV is strictly parallel or perpendicular to the jet because the Stokes $U$ is zero under integration both on the line of sight and along the cross-section of the jet projection.
Deviations from $U=0$ could be interpreted, for example, by the \textbf{B}-field disordering or introducing another additional parameter into the model. But \citet{LyutikovPG05} considered, like many other researchers, the velocity vector co-directional with the local axis of the jet.
The MOJAVE program results indicate the opposite \citep{Lister13,Lister16, Lister19, Lister21, Homan15,PushkarevKLS17}.
For one of the closest AGNs, M~87, using high angular resolution data, \citet{MertensLobanov16} found a rotational component in the jet flow motion.
Azimuthal component of the velocity vector is expected from theoretical models \citep[e.g., ][]{Hardee82, Beskin17}.
The model of the helical rotating jet \citep{But18a} has proven well when matching quasi-periods of long-term variability in the radio and optical ranges and interpreting changes in the inner jet position angle of the S5~0716+714 \citep{But18b} and OJ~287 \citep{ButP20} blazars. A similar curved jet, rotating around its axis, was also considered by \citet{VilRait99} for Mrk~501 and by \citet{Raiteri17} for CTA~102.
For the first time, we performed simulations under the assumption that the segment motion does not occur along its axis ($p\neq \rho$) and, moreover, the axis does not coincide with the radial direction.
Only in this case, the integral for Stokes $U$ over the line of sight will be different from zero. That results in a variety of angles of EV deviation from the local jet axis. Note that the integral Stokes $U$ over the line of sight is zero under $p=\rho>0^\circ$ too.
Distributions of the remaining polarization properties for the helical \textbf{B}-field are qualitatively consistent with the previously obtained simulation results in some cases. For example, as well as \citet{Murphy2013}, we reproduced: 1) parallel and perpendicular to the jet axis orientations of EVs or their combinations; 2) significant changes in PD; 3) one- and two-peak distributions of polarized intensity.
\citet{Murphy2013} concluded that a change of the twist angle by several degrees (from 41$^\circ$ to 53$^\circ$) is necessary to obtain different families of EV distributions. Our results indicate that changes in the twist angle of the field of several tens of degrees are needed to qualitatively change the transverse distribution of EVs (see \autoref{fig:heljetnonrad}, lines~4 and 8).
Although, in some sets of model parameters, different transverse distributions of EVs are realized depending on $\varphi$. 
Moreover, the nature of the EVs distribution varies depending on $p$ and $\rho$ \citep{But22}.
The dependence of the EV transverse distribution on both $\psi^\prime$ and the geometric and kinematic parameters of the jet segments does not allow us to uniquely determine the direction of the \textbf{B}-field twist based only on the EV distribution.
\autoref{fig:heljetnonrad} shows that longitudinal EVs on the axis and transverse EVs on the edge exist at the constant $\psi^\prime$, the value of which lies in a wide range. This fact is consistent with the conclusions of \citet{LyutikovPG05, Murphy2013, Clausen-BrownLyutikov11}, while \citet{Gabuzda21} emphasized a necessity of a change $\psi^\prime$ across the jet for such EV transverse distributions.

It is interesting to note the changes in $I$ and $P$ depending on the model parameters (see \autoref{fig:heljetnonrad}). For $p=2^\circ$, the distributions of $I$ at $\psi^\prime=45^\circ$ and $65^\circ$ are very similar and have a peak shift to the right side.
The corresponding distributions of $P$ have two peaks with different magnitudes. Finding the maximum peak on the left or right depends on $\psi^\prime$.
Comparing these distributions with those at $p=5^\circ$ and the corresponding $\psi^\prime$s, we can see that the dominant peaks change the side to the opposite. Thus, all transverse distributions of polarization properties depend not only on the angle and direction of the magnetic field twist but also on the geometrical and kinematic parameters of the jet segments. This fact indicates the necessity to compare simulations with observations simultaneously for all polarization parameters. 
In addition, to the helical \textbf{B}-field with the different twist angles, we examined the ``spine-sheath'' \textbf{B}-field configuration. Note that we did not initially associate it with the ``spine-sheath'' structure of an EV distribution, observed in some sources \citep{Pushkarev05} and which can be interpreted by the helical field. ``Spine-sheath'' \textbf{B}-field topology can occur in various ways. For example, the longitudinal field at the jet edge and the deceleration of the outflow outer layers can result from the jet's interaction with the environment \citep{Laing1980, Ghisellini05}. 
Alternatively, the Blandford-Znajek \citep{BlandfordZnajek77} and Blandford-Payne \citep{BlandfordPayne82} mechanisms may form spine and sheath, respectively. Different origination processes can result in a different twist of the magnetic field: almost toroidal in the spine and close to longitudinal at the jet edges, as, for example, for 0333+321 \citep{Asada08}.
Each of the scenarios has its observational evidence. Thus, the jet interaction with the ambient medium forming a shear layer is supported by magnetic field orientation, which is often found to be aligned with the outflow at its edges \cite[e.g.,][]{Attridge99, Pushkarev05,Croke10, MOJAVE_XXI}. The second scenario is favored by strong transverse velocity gradients found by \citet{MertensLobanov15} in the jet of M87. 
Analyzing the transverse gradient of the rotation measure, \citet{Gabuzda14} found evidence of a helical \textbf{B}-field in jets showing a change in the EV direction across the jet. Considering that the Faraday rotation mainly occurs in the external screen, this finding, together with our conclusion about the ``spine-sheath'' \textbf{B}-field topology, can indicate a \textbf{B}-field configuration similar to that in the ``cosmic battery'' model \citep[e.g., ][ and reference therein]{Contopoulos09}, but extending much farther beyond the boundaries of the detected jet.

We found a \text{good} qualitative correspondence, which takes place in all three considered \textbf{B}-field configurations. We emphasize that we considered a homogeneous and isotropic distribution of emitting electrons in the jet and constant jet velocity.
The exception is the case of the ``spine-sheath'' structure with the slow sheath, but there were no speed changes inside the spine and sheath.
On the other hand, \citet{Beskin17} analytically obtained a transverse jet distribution of the electron number density and flow velocity. Their use for the simulation of transverse polarization properties could affect some particular parts of the distributions. But due to the limited angular resolution, these features can be undetected in the VLBI observational data.
Noteworthy, under the simplest assumptions, namely, the homogeneous distribution of emitting electrons and the constant velocity of matter in the jet, all four observed transverse distributions, different for each considered object, are qualitatively well reproduced. 
We also note there is an insignificant discrepancy between two considered magnetic field topologies. Namely, the helical field with the small $\psi^\prime$ is similar to the case of the thick sheath. On the other hand, B-field with high $\psi^\prime$ resembles the case of the thin sheath. We cannot associate any topology with specific jet laughing mechanisms without a detailed analysis of the transverse gradient of the flow speed and jet interaction with the surrounding media.

\section{Conclusions}
\label{sec:conc}

We have simulated the transverse distributions of the linear polarization properties of parsec-scale AGN jets. We used various configurations of the strictly ordered magnetic field and accounted for the curved shape of jets and the non-radial motion of their segments. The main conclusions are as follows.

1) The Stokes U is zero only if the local jet axis coincides with the motion direction. In this case, the EV is perfectly either perpendicular or parallel to the jet. In the opposite case, deviations of EVs from the local jet axis lie in the range of 0$^\circ$ to 90$^\circ$.

2) Both the helical field and the ``spine-sheath'' structure can reproduce the basic forms of the observed transverse distributions. Namely, there are one- and two-peaked polarized intensity, U- and W-shaped PD distributions, longitudinal and transverse EV directions. At the same time, both quantitative and qualitative changes in the transverse distributions are possible only by changing the Doppler factor and/or the orientation relative to the observer of the jet segment.

3) Longitudinal EVs on one jet edge and transverse ones on the other can only be reproduced with the helical \textbf{B}-field. In this case, the angle between the \textbf{B}-field and the local jet axis in the reference frame of the source $\psi^\prime$ can be in a wide range of values. For the fixed both $\psi^\prime$ and rotation direction of the \textbf{B}-field, the positions of the jet sides containing longitudinal and transverse EVs also depend on the geometric and kinematic parameters of the outflow.

4) To determine the \textbf{B}-field configuration reliably, analysis of the distributions of all polarization parameters is necessary.

5) The model parameters, at which the observed and theoretical transverse distributions of total intensity and linear polarization properties agree for the 0333+321, 0836+710, and 1611+343 jets, are in narrow ranges of values. This indicates that the study of polarization is a powerful tool for probing and determining the physical, kinematic, and geometric parameters of the AGN jets.
The model, though, has difficulty reproducing one of the types of transverse profiles often observed in BL Lacertae objects -- the quasi-constant cuts of fractional polarization.

6) The obtained agreement between the model and observed transverse distributions of polarization properties indicates the well-ordered global magnetic field associated with parsec-scale AGN jets.

\section*{Acknowledgements}
We thank the anonymous referee for a careful reading of the article and useful comments, especially regarding the completeness of references and assistance in highlighting the main research results of the study.

This study was supported by the Russian Science Foundation: project 21-12-00241.
This research has made use of data from the MOJAVE database that is maintained by the MOJAVE team \citep{Lister18}.

\section*{Data Availability}

All simulated distributions are available in \url{ftp://jet.asc.rssi.ru/outgoing/pushkarev/transverse_cuts}. 



\bibliographystyle{mnras}
\bibliography{radiopolarization} 




\appendix

\section{Detailed description of the performed calculation}
\label{sec:appendix}

For an isotropic momentum distribution and a power-law energy distribution of emitting electrons $N(E)=K_e E^{-s}$ in the reference frame comoving with relativistic plasma, the Stokes parameters are \citep[see][]{LyutikovPG05}
\begin{equation}
    \begin{split}
        I=&\frac{s+7/3}{s+1} A \int d n \cdot |B^\prime \sin \chi^\prime|^{(s+1)/2}\,,\\
        Q=&A \int d n \cdot |B^\prime \sin \chi^\prime|^{(s+1)/2} \cos 2 \tilde \chi \,, \\
        U=&A \int d n \cdot |B^\prime \sin \chi^\prime|^{(s+1)/2} \sin 2 \tilde \chi\,,\\
        V=&0\,,
    \end{split}
    \label{eq:parstok}
\end{equation}
where
\begin{equation}
    A=\frac{\kappa (\nu)}{D_L^2 (1+z)^{2+(s-1)/2}} \cdot K_e \delta^{2+(s-1)/2}\,,
    \label{eq:A}
\end{equation}
where $D_L^2$ and $z$ are the luminosity distance and redshift of an object, $B^\prime$ and $\chi^\prime$ are the magnetic field strength and the angle of magnetic field direction with the line of sight in the comoving reference frame (which is denoted by a prime), respectively, $\tilde \chi$ is the observed angle between the wave electric vector and reference direction, 
\begin{equation}
   \delta=\{\Gamma[1-(\boldsymbol{\hat n},\boldsymbol{\beta})]\}^{-1}\,, 
    \label{eq:dopler}
\end{equation}
is the Doppler factor of the radiating jet segment, $\Gamma$ is the Lorentz factor, $\boldsymbol{\hat n}=\{\sin \theta_\rho,0,\cos \theta_\rho \}$ is the unit vector in the direction of emitting photons.  Here a hat over a vector means a unit vector in the given direction. In \autoref{eq:A}, the function $\kappa (\nu)$ is
\begin{equation}
\begin{split}
    \kappa(\nu)=\frac{\sqrt{3}}{4}\Gamma_\text{E}\left(\frac{3s-1}{12}\right)\Gamma_\text{E}\left(\frac{3s+7}{12}\right) \frac{e^3}{m_e c^2} \\ \left(\frac{3e}{2\pi m_e^3 c^5}\right)^{(s-1)/2} \nu^{-(s-1)/2}\,,
\end{split}
\end{equation}
where $e$ and $m_e$ are the charge and rest mass of an electron, respectively, $\Gamma_\text{E}$ is the Euler gamma function, and $\nu$ is the frequency of observations. We integrate the expressions in \autoref{eq:parstok} along the direction of the emitted photons (i.e., along the line of sight) over $d n= d x/ \sin \theta_\rho$ in the range from $-\sqrt{1-h^2}$ to $\sqrt{1-h^2}$, where $h$ is the distance between the line of sight and the projection of the jet axis onto the plane of the sky.

We assume a helical magnetic field with magnitude decreasing inversely proportional to the square of the distance from the local jet axis 
\begin{equation}
    \boldsymbol{B^\prime}=\frac{10}{1+\sqrt{x^2+y^2}} 
    \begin{Bmatrix} -\sin \psi^\prime \sin[\varphi(x,y)] \\
    \sin \psi^\prime \cos[\varphi(x,y)] \\
    \cos \psi^\prime \end{Bmatrix}\,,
\end{equation}
where $\varphi(x,y)$ is the azimuthal angle of the cylindrical coordinate system.

We find the angle $\chi^\prime$ from formula
\begin{equation}
    \cos \chi^\prime=\left(\boldsymbol{\hat{n}^\prime}, \boldsymbol{\hat{B}^\prime} \right)\,,
\end{equation}
where
\begin{equation}
    \boldsymbol{\hat n^\prime}=\frac{\boldsymbol{\hat n}+\Gamma\boldsymbol{\beta} \left[ \frac{\Gamma}{\Gamma+1}\left(\boldsymbol{\hat n},\boldsymbol{\beta} \right)-1\right]  }{\Gamma}\,.
\end{equation}

The angle $\tilde{\chi}$ is obtained from expressions 
\begin{equation}
    \begin{split}
        \sin \tilde \chi=\left( \boldsymbol{\hat e},\boldsymbol{l} \right)\,,\\
        \cos \tilde{\chi}=\left(\boldsymbol{\hat e}, \left[ \boldsymbol{\hat n},\boldsymbol{l}\right]  \right)\,,
    \end{split}
\end{equation}
where $\boldsymbol{l}=\{0,1,0\}$ is a vector orthogonal to the plane, containing $\boldsymbol{\hat n}$ and reference direction, $\boldsymbol{\hat e}$ is a unit vector in the direction of an electric vector in the wave, which is defined by equations
\begin{equation}
    \begin{split}
        \boldsymbol{\hat e}=&\frac{\left[\boldsymbol{\hat n},\boldsymbol{q^\prime} \right]}{\sqrt{q^{\prime 2}-\left(\boldsymbol{\hat n}, \boldsymbol{q^\prime} \right)^2}}\,,\\
        \boldsymbol{q^\prime}=&\boldsymbol{\hat{B}^\prime}+ \left[\boldsymbol{\hat n},\left[\boldsymbol{\beta}, \boldsymbol{\hat{B}^\prime}\right] \right]-\frac{\Gamma}{1+\Gamma} \left(\boldsymbol{\hat{B}^\prime},\boldsymbol{\beta} \right) \boldsymbol{\beta}\,.
    \end{split}
\end{equation}

\bsp	
\label{lastpage}
\end{document}